# Techno-Economic Analysis of Synthetic Fuel Production from Existing Nuclear Power Plants across the United States

Marisol Garrouste[a,b,d], Michael T. Craig[c], Daniel Wendt[a], Maria Herrera Diaz[a], William Jenson[a], Qian Zhang[a], Brendan Kochunas[b]
[a] Idaho National Laboratory
[b] University of Michigan, Nuclear Engineering & Radiological Sciences
[c] University of Michigan, School for Environment and Sustainability & Industrial and Operations Engineering
[d] Corresponding author, mgarrou@umich.edu

## 1  Keywords

## 2  Summary

Low carbon synfuel can displace transport fossil fuels such as diesel and jet fuel and help achieve the decarbonization of the transportation sector at a global scale, but large-scale cost-effective production facilities are needed. Meanwhile, nuclear power plants are closing due to economic difficulties: electricity prices are too low and variable to cover their operational costs. Using existing nuclear power plants to produce synfuels might prevent loss of these low-carbon assets while producing synfuels at scale, but no technoeconomic analysis of this Integrated Energy System exist. We quantify the technoeconomic potential of coupling a synthetic fuel production process with five example nuclear power plants across the U.S. to explore the influence of different electricity markets, access to carbon dioxide sources, and fuel markets. Coupling synfuel production increases nuclear plant profitability by up to $792 million in addition to a 10% rate of return on investment over a 20 year period. Our analysis identifies drivers for the economic profitability of the synfuel IES. The hydrogen production tax credit from the 2022 Inflation Reduction Act is essential to its overall profitability representing on average 75% of its revenues. The carbon feedstock transportation is the highest cost – 35% on average – closely followed by the synfuel production process capital costs. Those results show the key role of incentive policies for the decarbonization of the transportation sector and the economic importance of the geographic location of Integrated Energy Systems.

# 1. Introduction

Transportation is a difficult-to-decarbonize sector of the global economy. Decarbonization strategies for this sector often focus on vehicle electrification. However, electrification strategies for aviation[1], marine[2], and heavy ground transportation[3] present cost, range, infrastructure, and power supply challenges[1–5]. These applications currently rely on the higher energy density of liquid fuel, and a large-scale overhaul of the supporting infrastructure and fleet operations remains a challenge. For all kinds of transportation, the liquid fuel infrastructure is one of the most developed networks of resources. Low-carbon synthetic fuel – synfuels - thus represent a promising decarbonization pathway for the transportation sector as they can displace liquid fossil fuels without radical changes to the fuel distribution and use infrastructure[6]. Studies emphasize the crucial role of synfuels in clean energy transitions[7,8]. In 2022, the $CO_2$ emissions from the global transportation were nearly 8 Gt $CO_2$[9]. With carbon from atmospheric sources and a low-carbon energy such as nuclear, the reduction in greenhouse gases emissions could as high as 92% for diesel and 50% for jet fuel when replacing conventional liquid fossil fuels with synthetic fuels[10]. Trucks and buses combined emitted 2.3 Gt CO2 in 2022, replacing the diesel fuel used by these vehicles with this alternative low-carbon synthetic counterpart could avoid 2.1 Gt CO2. For the aviation sector the avoided emissions would be 0.4 Gt CO2. With the displacement of conventional petroleum-based diesel and jet fuel by their synthetic low carbon alternatives from nuclear energy, the transportation sector could see its emissions reduced by more than 30%. About three quarters of the reductions needed to abide by the Net Zero Emissions by 2050 scenario could thus come from synthetic diesel and jet fuel alone[9].

Various low-carbon energy sources (including renewables, natural gas with carbon capture and storage, and nuclear energy) powering various synfuel production pathways (including biomass-to-liquid, Fischer Tropsch (FT), coal-to-liquid, gas-to-liquid and hydrogenation to methanol) have been considered[11–17.] To produce low-carbon synfuel, carbon feedstock must ultimately come from the atmosphere, reducing the choice of potential carbon feedstocks to either biomass or $CO_2$ from waste-streams that would otherwise be released into the atmosphere. Comparison between synfuel production processes have shown that biomass to liquid technologies are still at the development stage and for coal or gas to liquid technologies, economies of scale can enable large facilities to compete with conventional fossil fuels[11]. Solar PV has been considered to power a methanol production facility using a biogas feedstock[12], and to produce synthetic fuel when integrated into the European grid[13]. The decarbonization of the transportation sector using nuclear energy has only been considered with biomass as a feedstock: such a strategy could compete with fossil fuels but would require a large-scale agricultural and logistical effort[14]. A comparative review of $CO_2$ conversion technologies presents performance indicators for the Fischer-Tropsch process and hydrogenation to methanol technology, the two carbon dioxide ($CO_2$) conversion technologies leading to the production of drop-in liquid fuels[15]. While the FT process presents higher investment costs, the operational and maintenance costs are higher for the hydrogenation to methanol technology. The technology readiness level of those technologies indicates that the methanol pathway may soon be ready for commercialization while the FT has already been used in commercial operations. Finally, while the FT process has a higher electricity usage it also presents a higher $CO_2$ utilization rate and leads to higher value products indicating a higher economic and environmental potential. Other studies corroborate these findings[16,17]. They find that the methanol and FT pathways are the most promising from a technical and economic standpoint, the methanol pathway being right behind the FT pathway in terms of technology readiness and economic potential of the final products.

At the international level, the nuclear industry has experienced renewed interest as recent geopolitical events have shown the vital importance of energy independence supported by nuclear reactors[18]. However, in Northern America, the nuclear industry is struggling. In its most recent report on Energy, Electricity and Nuclear, the IAEA highlights the role of nuclear power in $CO_2$ emissions reductions[19] but in their low scenario the U.S. nuclear capacity should reduce by 2030. Across the U.S., nuclear power plants (NPP) are shutting down. Due to operational losses[10], utilities owning them are losing money by continuing to run them. Without any help to regain profitability nuclear power plants in the U.S. may retire prematurely, depriving the country of a large capacity of low-carbon energy[20]. Proposed solutions to increase the economic viability of NPPs include direct subsidies corresponding to their positive externalities[20], or the diversification of their production with hydrogen generation[21,22]. Given their large capacity, reliability, and availability, NPPs would be prime candidates to power decarbonized synthetic fuel production processes, which require a large and steady inflow of electricity and heat. Producing synfuel via the FT pathway can benefit from economies of scale[15] and thus the current fleet of NPPs could efficiently produce the necessary heat and electricity needs for this process. The coupling of a LR and a FT synfuel production process into an Integrated Energy System (IES) would offer new economic opportunities for the NPP by strategically expanding the range of products: synthetic fuels prices do not suffer from ample hourly variations, and they tend to be higher than electricity[23]. Such a strategy has only been considered from a theoretical perspective[24].

Despite its economic and technical potential for decarbonizing the transportation sector, no techno-economic analysis has been performed for such a synfuel-nuclear IES. To assess its feasibility, we answer three research questions: What is the profitability of a synfuel IES versus a standalone NPP? What are the key drivers for the economic profitability of a synfuel IES? What is the global decarbonization potential of displacing conventional fossil fuels with synfuel via this type of synfuel IES? To capture variability in electricity markets, feedstock costs, and NPP characteristics, we run our analysis for five NPPs across the United States: Davis-Besse (Ohio), Braidwood (Illinois), South Texas Project (Texas), Cooper (Nebraska), and Prairie-Island (Minnesota). For each NPP, we optimize the IES configuration and operations to maximize profitability. The synfuel IES (Figure 1) configuration includes several coupled components: the NPP produces power for the grid or the High-Temperature Steam Electrolysis plant (HTSE), which also consumes heat accounted for as electricity; hydrogen produced by the HTSE is either stored or sent to the Fischer-Tropsch process (FT); and the FT process produces a mix of synthetic fuel products (diesel, jet fuel, and naphtha) from $CO_2$ and hydrogen. To capture the effect of short- and long-term variability in electricity prices on profitability, we optimize operations at an hourly interval using over 10 years of historical data. Finally, to assess the robustness of our results against uncertainties in model inputs, we perform a sensitivity analysis, ranking and quantifying the influence of these uncertainties.

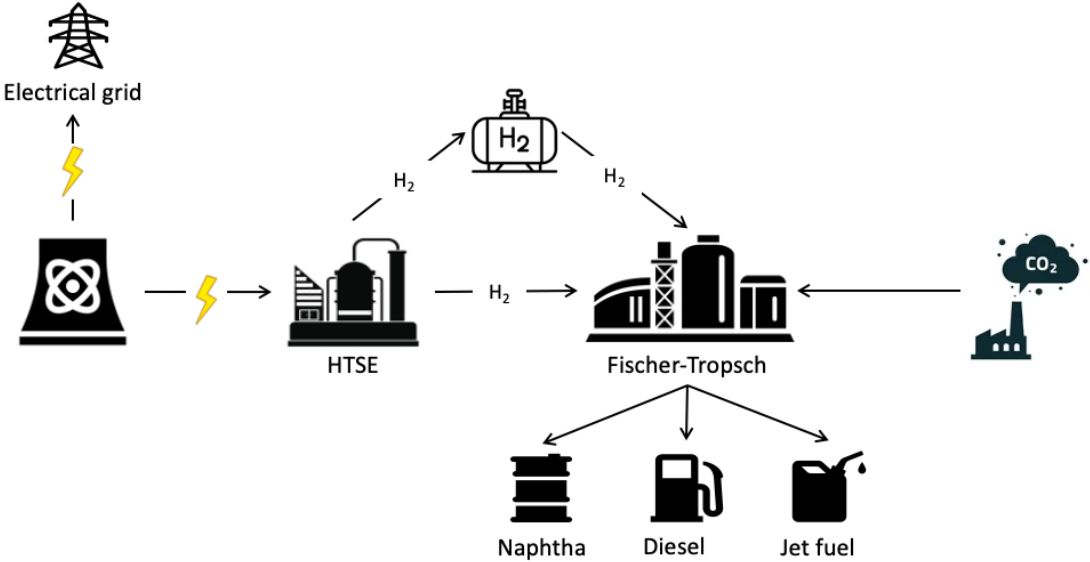

*Figure 1 Synfuel IES: grid-integrated synfuel production process via nuclear energy.*

## 2. Materials and Methods

### 2.1 Power System Modeling via HERON

To model the synfuel IES presented in Figure 1 we used HERON[25] (Holistic Energy Resource Optimization Network) and RAVEN[26] (Risk Analysis Virtual ENvironment). Using these codes, we can model the synfuel IES within the electricity market context and optimize the system's configuration on average taking price uncertainties into account. We give an overview of the HERON code, of the ARMA (Auto-Regressive Moving Average) models used to represent stochastic electricity prices and we detail the synfuel IES' modeling and optimization.

#### 2.1.1 HERON overview

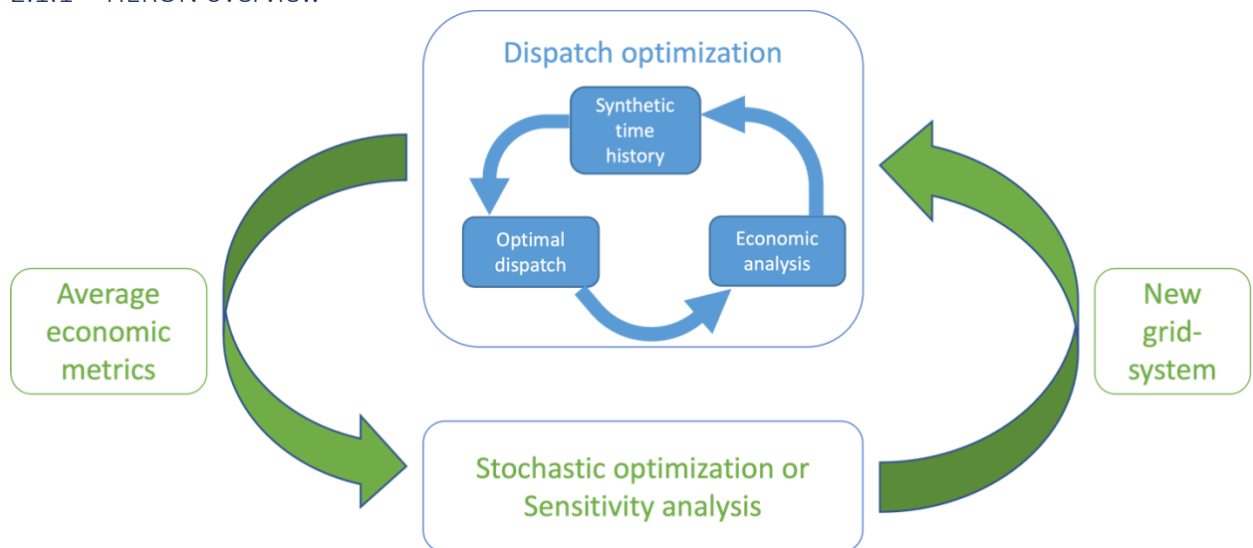

*Figure 2 HERON stochastic power system optimization[22]*

Figure 2 shows the HERON stochastic power system optimization process used to determine the optimal configuration for an IES to maximize its NPV. We explain it in the context of the synfuel IES. The goal of this process is to find the optimal set of capacities for the HTSE, the $H_2$ storage, and the FT process given the NPP capacity at the considered location. HERON can be run in either optimization or sensitivity analysis mode. In optimization mode, the simulation will result in a single final point corresponding to the system's configuration maximizing the chosen econometric. In sensitivity analysis mode – the mode used for the present analysis – the user provides a set of capacities to evaluate for each component part of the analysis. All combinations of these capacities are then simulated, and a better understanding of the system's response surface can be acquired. An optimization or sensitivity loop point consists in a capacity value for the HTSE, FT and $H_2$ storage. The system with fixed components' capacities is then analyzed within the dispatch optimization loop. The dispatch optimization loop consists of solving the optimal dispatch problem over the plant lifetime. To account for the stochasticity of electricity prices, it is solved multiple times with different realizations of electricity price profiles from the ARMA model trained for the location. The average NPV and its standard deviation are computed and sent back to the optimization or sensitivity loop.

### 2.1.2 ARMA stochastic synthetic time series

Electricity market prices are unpredictable and among the main drivers of the optimal resource utilization in a grid-integrated energy system. Using the RAVEN signal analysis capabilities[27,28] we generate independent synthetic time series with consistent statistical behavior. The initial data is detrended via a Fourier analysis to identify potential periodicity and an Auto-Regressive Moving Average - ARMA - is fitted on the residual noisy data (more information on the ARMA model is available in SI Section Synthetic Time Series via the Auto-Regressive Moving Average model). For each location in our analysis, up to 10 years of hourly Day-Ahead Market (DAM) electricity prices data (those distributions are shown on Figure 3) are collected and an ARMA model is trained on this historical data. We validate the ARMA model representativity by comparing statistical moments of the historical and synthetic electricity prices distribution, this validation is presented in SI section ARMA Model Validation.

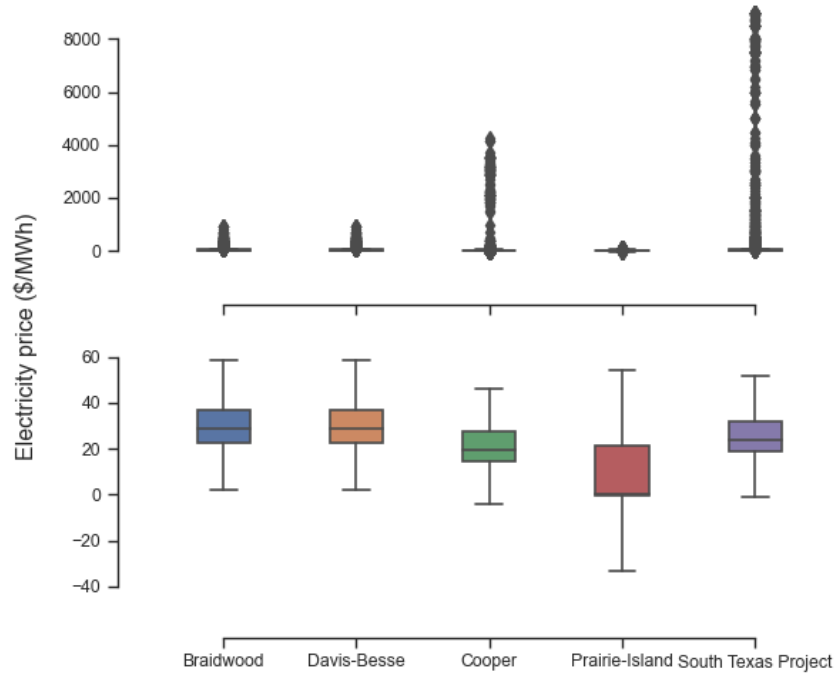

*Figure 3 Historical electricity prices distributions at each location.*

### 2.1.3 Synfuel IES HERON modeling

Figure 1 shows the synfuel IES as modelled in HERON. The High-Temperature Steam Electrolysis (HTSE), $H_2$ storage, and Fischer-Tropsch (FT) process are components whose capacity is optimized. The primary source of energy is the NPP, which produces electricity for the grid and electricity and heat for the synfuel production process. NPPs present different capacities depending on the location (Table 1). The heat coupling with the HTSE is modelled as an electrical coupling assuming a 33% conversion efficiency. The portion of electricity not consumed by the HTSE is sent to the electricity market the NPP is located in. We assume electricity prices are not affected by the electricity production of the NPP, a common assumption in single generator analysis. Pricing data is also obtained from an Auto-Regressive Moving Average – ARMA - model trained on historical data and reproducing its periodical patterns and variability.

The synfuel production process starts with the HTSE plant flexibly producing hydrogen. The HTSE operates in hot-standby mode. Given an additional consumption of heat and electricity, it can shut down and restart in less than an hour, the modelling time step considered in this analysis. Costs and process performance data were obtained by modelling this plant in ASPEN[29]: the HTSE has a specific electricity and thermal energy requirements of 36.8 kWh-e/kg-H2 and 6.4 kWh-t/kg-H2 and a total capital cost of $703/kW-DC and fixed variable costs respectively amount to $32.6 MW-DC/year and 3.4 MWh-DC.

The hydrogen produced by the HTSE is either sent to a storage facility or the FT process. The FT process operates at steady state with constant feedstock inflow and product outflow rates to maximize efficiency, a common operational strategy in refining processes. The hydrogen storage facility is there to ensure this constraint is met whenever the hydrogen production from the HTSE decreases. We assume a CAPEX cost of $500/kg for the hydrogen storage facility from a review of storage costs[22]. The FT takes in $CO_2$ as an inflow from surrounding $CO_2$ emitting facilities. Costs and efficiency parameters were obtained from an ASPEN modelization of this process[30]. For one kilogram of hydrogen consumed the FT process produces 0.69 kg of naphtha, 0.84 kg of jet fuel and 0.46 kg of diesel. The total capital costs amount to 158,102,945 $ for a plant with a 10,625 kg-$H_2$/h capacity and fixed and variable O&M costs to $7,640,007/(kg-$H_2$/h) and $21,732,221/(kg-$H_2$/h).

We model the BAU – Business-As-Usual – case in HERON to compute the Δ(NPV) econometric. The current use of NPPs is represented by the BAU case: NPPs produce electricity for the grid as a baseload power supplier. We assume that no load-following strategy is in place: NPPs are price-takers – a common assumption in single generator analysis, they sell their electrical production at the Day-Ahead Market (DAM) rate.

| NPP | Reactor Type | Capacity | Electricity Market | State | Synfuel Region |
|---|---|---|---|---|---|
| Braidwood (Unit 1) | PWR | 3645 MWth 1194 MWe | PJM | Illinois | East North Central–IL |
| Davis-Besse | PWR | 2817 MWth 894 MWe | PJM | Ohio | East North Central–OH |
| South Texas Project (Unit 1) | PWR | 3853 MWth 1280 MWe | ERCOT | Texas | West South Central |
| Prairie Island (Unit 1) | PWR | 1677 MWth 522 MWe | MISO | Minnesota | West North Central–MN |
| Cooper Nuclear Station | BWR | 2419 MWth 769 MWe | SPP | Nebraska | West North Central–NE |

Table 1 NPPs characteristics at each location.

| Parameter | Value | Description |
|---|---|---|
| Project life | 20 years | Representative of NRC operating license extension for existing NPPs |
| Weighted average cost of capital (WACC) | 10% | Cost of capital, minimum return to be earned on assets to satisfy investors |
| Inflation rate | 2.18% | Average for 2000–2021 |
| Federal corporate tax rate | 21% | |
| State corporate tax rate | - | Value depends on location of NPP |
| Depreciation schedule | 15 years MACRS | |
| Synthetic fuel price | - | Regional, see section 2.3 |

Table 2 Synfuel IES HERON model financing parameters.

### 2.1.4 Configuration selection at each location

One thousand configurations were run for each location to map the decision variables space that correspond to the capacities of the components of the synfuel production process: the HTSE, the FT and the hydrogen storage. This approach - running HERON in sensitivity analysis mode - was chosen to overcome the shortcomings of a classical optimization of the components' capacities as it allows for better understanding of the response surface of the system. For each IES component, ten equidistant points for the range of capacities possible for each component were analyzed. The FT and HTSE process were modelled[21] from 100 to 1000 MWe of total power requirement. The FT process has a constant electricity demand of 14.9 MWe[24]. The HTSE capacity cannot be larger than the capacity of the NPP since we define our system as an Integrated Energy System with no outside energy input. Consequently, the maximum capacity for the HTSE corresponds to the minimum among 985.1 MWe and the NPP capacity and the maximum for the FT capacity corresponds to the maximum hydrogen output flow from the HTSE. For the hydrogen storage capacity range, we calculated the range of capacities to be tested based on the maximum capacity of the FT process since the storage exist to ensure a steady input flow of hydrogen to the FT process. We tested a range of storage capacity from no storage to a capacity corresponding to a day's worth of the maximum capacity FT process consumption. Table 3 shows a summary of the range of the components' capacities selected through these methods.

| Plant | Capacity (MWe) | HTSE (MWe) | | FT (ton-H2/h) | | Storage (ton-H2) | |
|---|---|---|---|---|---|---|---|
| | | Min | Max | Min | Max | Min | Max |
| Braidwood | 1194 | 85.1 | 985.1 | 2.14 | 24.8 | 0 | 594.1 |
| Cooper | 769 | 85.1 | 754.1 | 2.14 | 19.0 | 0 | 454.8 |
| Davis-Besse | 894 | 85.1 | 879.1 | 2.14 | 22.1 | 0 | 530.2 |
| Prairie-Island | 522 | 85.1 | 507.1 | 2.14 | 12.7 | 0 | 305.8 |
| South Texas Project | 1280 | 85.1 | 985.1 | 2.14 | 24.8 | 0 | 594.1 |

Table 3 Range of components capacities.

### 2.1.5 Economic Analysis

To compare the profitability of the synfuel IES and the BAU use of reactors we calculate the differential NPV as presented in the following equation:

$$\Delta NPV = NPV \text{ (Synfuel IES)} - NPV(BAU)$$

The computation of the NPV is performed in HERON and corresponds to the difference between the net cash inflows and outflows over the plant lifetime. By calculating differences in NPVs, redundant cashflows between the two systems being compared cancel out. We assume that the NPPs always run at 100% capacity. For each location the optimal configuration corresponds to the one that maximizes the Δ(NPV). This metric is a linear function of the NPV of the synfuel IES, so the optimal IES configuration is the IES configuration with the highest NPV.

### 2.1.6 Sensitivity Analysis

A sensitivity analysis is performed to assess the robustness of the results against uncertainties and variations in input parameters. The selected input parameter is perturbed by a fixed amount or percentage with every other parameter staying constant. The HERON model of the synfuel IES is then re-run with the optimized configuration for the corresponding location.

The influence of each parameter is assessed via the change in profitability corresponding to the variation in Δ(NPV) obtained after perturbing the input parameter. It is calculated as follows:

$$\text{Change in profitability (\%)} = \frac{\Delta(NPV)_{case} - \Delta(NPV)_{ref}}{\Delta(NPV)_{ref}} = \frac{NPV_{case} - NPV_{ref}}{NPV_{ref} - NPV_{baseline}}$$

Where the $NPV_{case}$ is the NPV of the perturbed case to evaluate, $NPV_{ref}$ is the NPV of the optimized synfuel IES, and $NPV_{baseline}$ is the NPV of the BAU case.
The rationale beyond the perturbation of the input parameters is explained in Table 4.

| Input parameter | Perturbed value | Description/Rationale |
|---|---|---|
| $H_2$ PTC | $0/kg | To evaluate the profitability of the synfuel IES without this subsidy |
| | $1/kg | Corresponds to the third tier of the clean hydrogen production tax credit[31] |
| | $2.7/kg | If the utility owning the synfuel IES cannot directly benefit from tax credit it can sell it for 90% of its value[32] |
| Synfuels prices | ± 25% | Adjustment to the synfuel prices projection in the EIA Annual Energy Outlook |
| CAPEX | ± 25%. | For every component in the synfuel production process (HTSE, FT, and $H_2$ storage) |
| $CO_2$ feedstock cost | $30/ton $60/ton | In the reference case we assume this feedstock is free and is a waste stream from other industries that must be transported to the NPP location via pipelines. For the sensitivity analysis we consider an additional cost of $30/ton and $60/ton. The value of $60/ton correspond to the one that could be credited to a $CO_2$ emitter under Section 45Q of the U.S. Internal Revenue Code for using the $CO_2$ for an industrial use. The availability of this credit may increase the market value of the $CO_2$ emitted from industrial sources |
| O&M costs | ± 25% | For the components of the synfuel production process |

*Table 4 Perturbation of the input parameters for the sensitivity analysis.*

## 2.2 CO$_2$ feedstock transportation cost

CO$_2$ is needed to produce synfuel via the FT process. The amount of CO$_2$ feedstock needed depends on the capacity of the synfuel plant at each location. An upper bound on the CO$_2$ demand was calculated for each location considering the capacity of the NPP. CO$_2$ sources across the U.S. were mapped using reports from the Environmental Protection Agency (EPA)[33]. Seven sources of carbon were considered: bioethanol, ammonia, natural gas (power plant), coal (power plant), hydrogen, steel, and cement plants. For each type of source, the availability and the capture cost vary. The capture and transportation costs via pipeline to use the CO$_2$ from a source at the NPP location were computed via a model developed by the National Energy Technology Laboratory (NETL)[34]. For each location, supply curves were developed to estimate the cost of the CO2 feedstock as a function of the amount required. These supply curves are shown in Appendices section CO$_2$ feedstock transportation cost model on Figure 10**Error! Reference source not found.** that present the full CO$_2$ supply curve development process.

## 2.3 Synfuel market analysis

To quantify revenues from synfuel sales, we forecast refinery gate prices at which fuel produced by the synfuel IES would be sold. NPPs included in this analysis are located in regions defined by the U.S. Census Bureau[35] ('Synfuel Region' column of Table 5). For each region, the U.S. Energy Information Administration (EIA) provides fuel prices forecasts with yearly price points, we used the 2022 EIA Annual Energy Outlook forecasts. These were adjusted by removing applicable state and federal fuel taxes as well as transportation, marketing and distribution costs[36]. Marketing and distribution costs were estimated using reports from the EIA[37] and transportation costs were estimated based on the likely transportation methods used for each product given the location of the NPP. Current tax rates are used. As forecasts for the price of naphtha were not readily available but the historical correlation between the gasoline and naphtha prices is high[36], gasoline prices' forecasts were used and adjusted to produce forecasts for the naphtha prices. Table 5presents the resulting fuel products prices for the four census fuel regions included in our analysis over the 2022-2050 period. We refer the reader to SI Section

|          | Braidwood  |           | Cooper     |           | Davis-Besse |           | Prairie-Island |           | South Texas Project |           |
|----------|------------|-----------|------------|-----------|-------------|-----------|----------------|-----------|---------------------|-----------|
|          | Historical | Synthetic | Historical | Synthetic | Historical  | Synthetic | Historical     | Synthetic | Historical          | Synthetic |
| mean     | 33.0       | 33.0      | 25.8       | 25.5      | 33.0        | 33.0      | 9.4            | 9.4       | 41.2                | 41.6      |
| std      | 23.1       | 23.1      | 110.3      | 110.6     | 23.1        | 23.1      | 15.1           | 15.1      | 290.7               | 290.5     |
| min      | 2.3        | 2.3       | -65.1      | -65.1     | 2.3         | 2.3       | -66.2          | -66.2     | -20.2               | -20.2     |
| 25%      | 22.7       | 22.6      | 14.8       | 14.4      | 22.7        | 22.6      | -0.5           | -0.6      | 18.9                | 16.8      |
| 50%      | 28.8       | 28.8      | 19.3       | 19.2      | 28.8        | 28.8      | 0.5            | 0.6       | 23.9                | 24.4      |
| 75%      | 37.0       | 37.1      | 27.2       | 27.2      | 37.0        | 37.1      | 21.5           | 21.5      | 32.0                | 36.4      |
| max      | 933.7      | 933.7     | 4231.0     | 4231.0    | 933.7       | 933.7     | 97.0           | 97.0      | 8996.8              | 8996.8    |
| kurtosis | 243.3      | 241.8     | 755.9      | 750.0     | 243.3       | 241.2     | 1.2            | 1.2       | 699.6               | 696.7     |
| skewness | 10.9       | 10.8      | 26.8       | 26.6      | 10.9        | 10.8      | 1.3            | 1.3       | 25.8                | 25.7      |

*Table 10 Synthetic and historical electricity prices distribution comparison for the validation of ARMA models.*

To perform the validation of the ARMA model trained on historical electricity market data we compute and compare statistical moments for the historical and synthetic data at each location (Table 10). The synthetic data should reproduce the historical data and we observe that the median, quartiles, minimum and maximum are identical. The kurtosis and the skewness of the distribution are close. They are respectively a measure of the importance of outliers in the distribution and of the symmetry of the distribution and these characteristics of the electricity prices distribution will influence the behavior of the synfuel IES as extreme electricity prices will change the relative value of the synfuel products compared to electricity.

Synfuel Market and Prices Forecasts for more detailed information on how the forecast for fuel prices were calculated.

|  | Average price ($/gal) (2022-2050) | | | | |
|---|---|---|---|---|---|
| Product | West North Central - NE | West North Central – MN | West South Central | East North Central - OH | East North Central - IL |
| Naphtha | 0.75 | 1.08 | 0.71 | 1.15 | 0.82 |
| Diesel | 1.83 | 1.80 | 1.99 | 1.79 | 1.55 |
| Jet Fuel | 1.93 | 1.80 | 2.00 | 1.88 | 1.85 |

*Table 5 Average fuel product price among the four fuel regions for the 2021-2050 period.*

# 3   Results

## 3.1   Economic profitability of the synfuel IES

Table 6 presents a summary of the optimal configuration selected for each location to maximize the economic profitability of the synfuel IES. In general, the optimal IES configuration includes a large HTSE relative to the electrical output of the NPP. HTSE sizes range from 507 to 985 MWe, or 83% to 98% of each NPP output. FT production is sized to match the output of the HTSE process at two out of five locations, and at 90% of HTSE output for the others. Maximizing the capacity of the FT is profitable since a $3/kg hydrogen Production Tax Credit (PTC) is collected when the low carbon hydrogen produced by the HTSE is used in the FT process, which we further examine below. Optimal IES configurations have small or no hydrogen storage. Three IES deploy 34-66 tons of hydrogen storage, equivalent to about 2.7 hours of HTSE output. This small storage capacity allows the synfuel IES to continue producing synfuels at a constant rate while sending power to the grid instead of the HTSE during high-prices hours. The FT process is operated in a steady-state manner, a common operational strategy in refining operations.

|  | Existing NPP capacity and optimized synfuel IES capacities | | | | |
|---|---|---|---|---|---|
| Location | NPP (MWe) | HTSE (MWe) | HTSE (ton-$H_2$/h) | FT (ton-$H_2$/h) | $H_2$ storage (ton-$H_2$) |
| Braidwood | 1193 | 985 | 24.8 | 22.2 | 66.0 |
| Cooper | 769 | 680 | 17.1 | 17.1 | 0.0 |
| Davis-Besse | 894 | 791 | 19.9 | 19.9 | 0.0 |
| Prairie-Island | 522 | 507 | 12.7 | 11.6 | 34.0 |
| South Texas Project | 1280 | 985 | 24.8 | 22.2 | 66.0 |

*Table 6 Optimal configuration for the synfuel IES at each location.*

Figure 4 shows the added profitability of the synfuel IES compared to the current use of each NPP, or only selling electricity to the grid, referred to as the Business-As-Usual case (BAU). At each location, the differential Net Present Value ($\Delta$(NPV)) is calculated over a 20 year period, corresponding to the license extension granted by the NRC to existing NPPs. It measures the additional revenue from the synfuel IES compared to the BAU case beyond an assumed 10% rate of return on investment for the synfuel production process. The added profitability ranges from -$336M to $792M USD (2020). The synfuel IES is more profitable than the BAU strategy at four out of five locations; only at the South Texas Project location would it be not profitable to invest in low carbon synfuel production. The variability of electricity prices insert uncertainty in the NPV of the systems considered resulting in up to 80% uncertainty in the ($\Delta$(NPV).

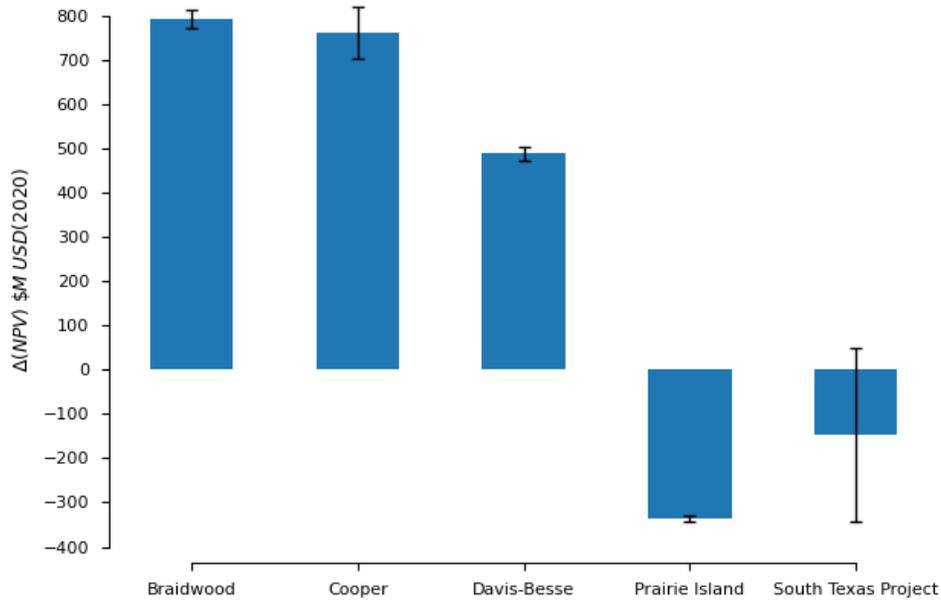

*Figure 4 Additional profitability (Δ(NPV) over 20 years) at each location for the optimal synfuel IES configuration. Error bars indicate the 95% confidence interval stemming from the stochasticity in electricity prices.*

We answer our second research question about the identification of the key drivers of the economics of the synfuel IES by examining the normalized cashflows (by the NPP capacity) for each location (Figure 5). A blue horizontal line indicates the NPV of the BAU case, reflecting the electricity price distribution and capacity for each NPP. For instance, the prices at the South Texas Project location are higher than the ones at the Prairie Island location and lead to more than a fourfold difference in the NPV of the BAU case. We observe that the hydrogen PTC dominate IES revenues and is thus critical to ensure the profitability of the synfuel IES at each location. Hydrogen PTC revenues account for 64 to 75% of total IES revenues. The combined revenues from synfuel products sales are the next largest driver of IES revenues, representing on average 25% of the revenues. Finally, revenues from electricity sales vary between locations. At the Davis-Besse and Prairie-Island locations the optimal configuration for the synfuel IES corresponds to a synfuel dedicated system and thus no electricity is sent to the grid. At other locations electricity sales represent 3 to 14% of the revenues. At each location we note the existence of threshold electricity prices above which the proportion of the NPP electricity production sent to the grid sharply increases (see SI section The grid stabilizer role of the NPP within the synfuel IES).

The $CO_2$ feedstock cost and the CAPEX represent the largest costs: the $CO_2$ feedstock cost is particularly high at the South Texas Project location while the CAPEX costs dominate at the Braidwood location where the $CO_2$ costs are lower thanks to a closer proximity to high-quality carbon sources reflected by the supply curve depicted in Appendices section $CO_2$ feedstock transportation cost model (Figure 10). O&M costs represent on average 26% of total costs, while foregone capacity market payments represent 8 to 10% of the costs.

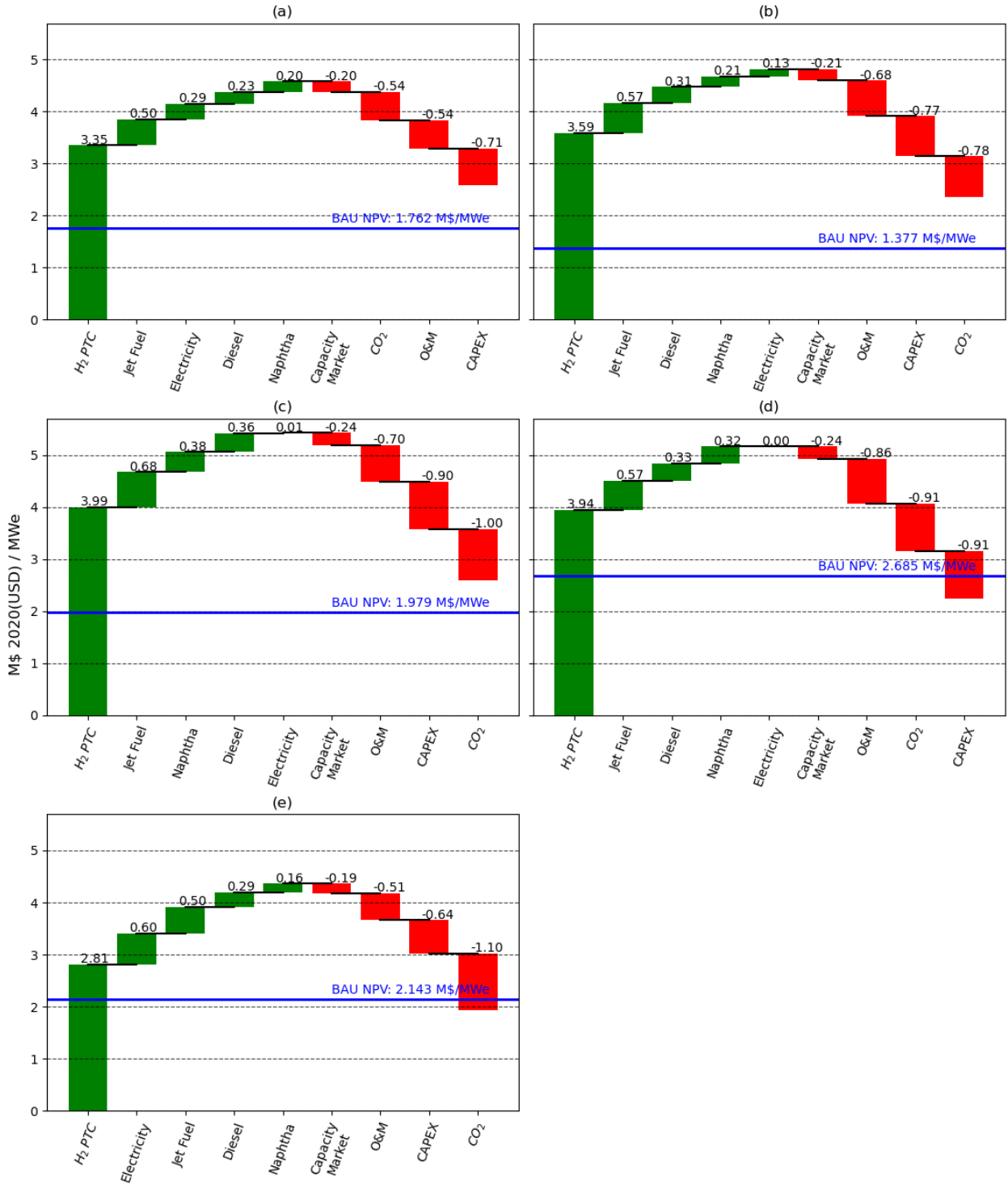

*Figure 5 Cashflows breakdown for the Braidwood location (a), the Cooper location (b), the Davis-Besse location (c), the Prairie-Island location (d) and the South Texas Project location (e).*

## 3.2 Sensitivity analysis

We test the robustness of our results against each system's configuration variations and against the uncertainties in our input parameters at each location. Only the $CO_2$ feedstock cost and the $H_2$ PTC value could decrease the profitability of the synfuel IES and make it less interesting than the BAU NPP as shown on Figure 6 (see the Experimental Procedures section for the definition of the change in profitability). The value of the hydrogen PTC has the largest influence on the economics of the synfuel IES. A decrease of 10% of its value ($3 to $2.7/kg-H2) induces a 20% to 125% decrease in the profitability of the synfuel IES depending on the location. Without any hydrogen PTC, no IES are more profitable than the BAU NPP selling electricity to the grid. Additional costs for the $CO_2$ feedstock would have a major influence on the economics of the synfuel IES: at a $60 per ton cost – an additional corresponding to the credit given to $CO_2$ emitter under Section 45Q of the U.S. Internal Revenue Code for an industrial use - the profitability would drop by about 100% at three out of five locations. The CAPEX, O&M costs, prices of the synfuel products and electricity prices only have a significant influence at the South Texas Project location but little at the other locations, we refer the reader to SI Section Sensitivity Analysis Results for the corresponding figures. At this location, the electricity prices are highly variable, and the cost of the $CO_2$ feedstock is highest, so this location is particularly sensitive to the input perturbations applied. For instance, a 25% increase in electricity prices would increase the profitability of the synfuel IES by 100% making it financially profitable to invest in the synfuel production process.

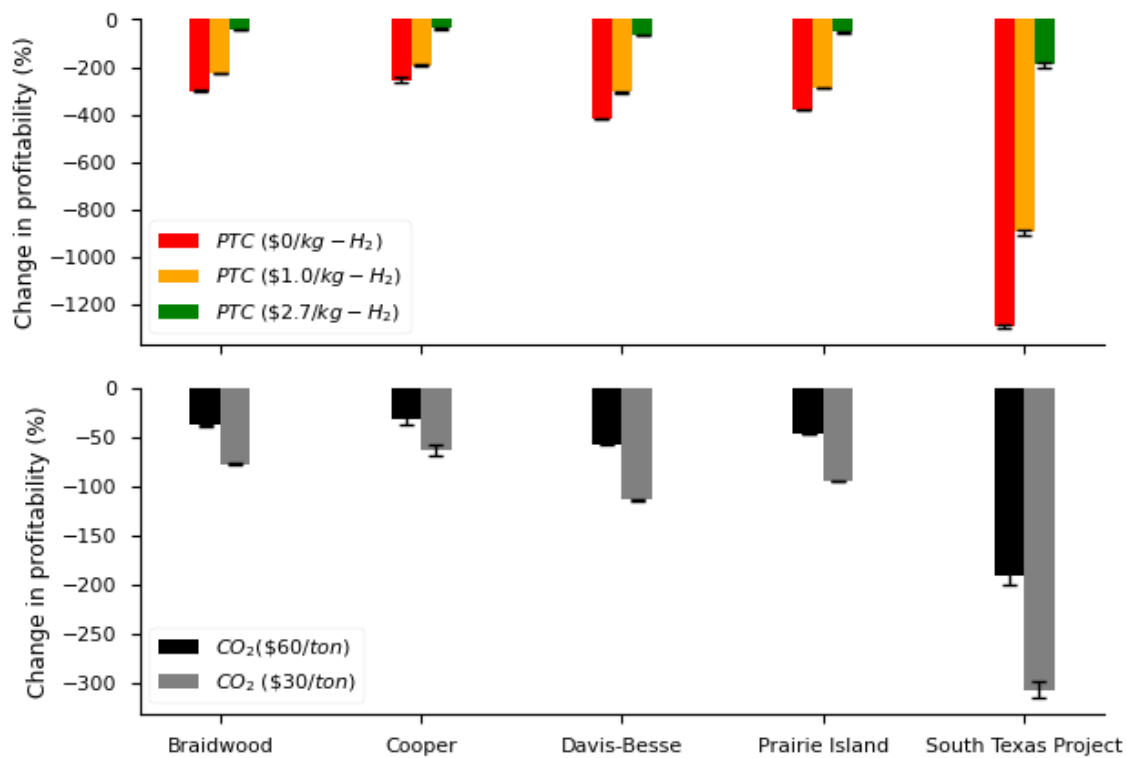

*Figure 6 The value of the hydrogen PTC and the cost of the $CO_2$ feedstock are the most influential parameters, Sensitivity analysis results for the carbon feedstock cost and the hydrogen PTC value at each location.*

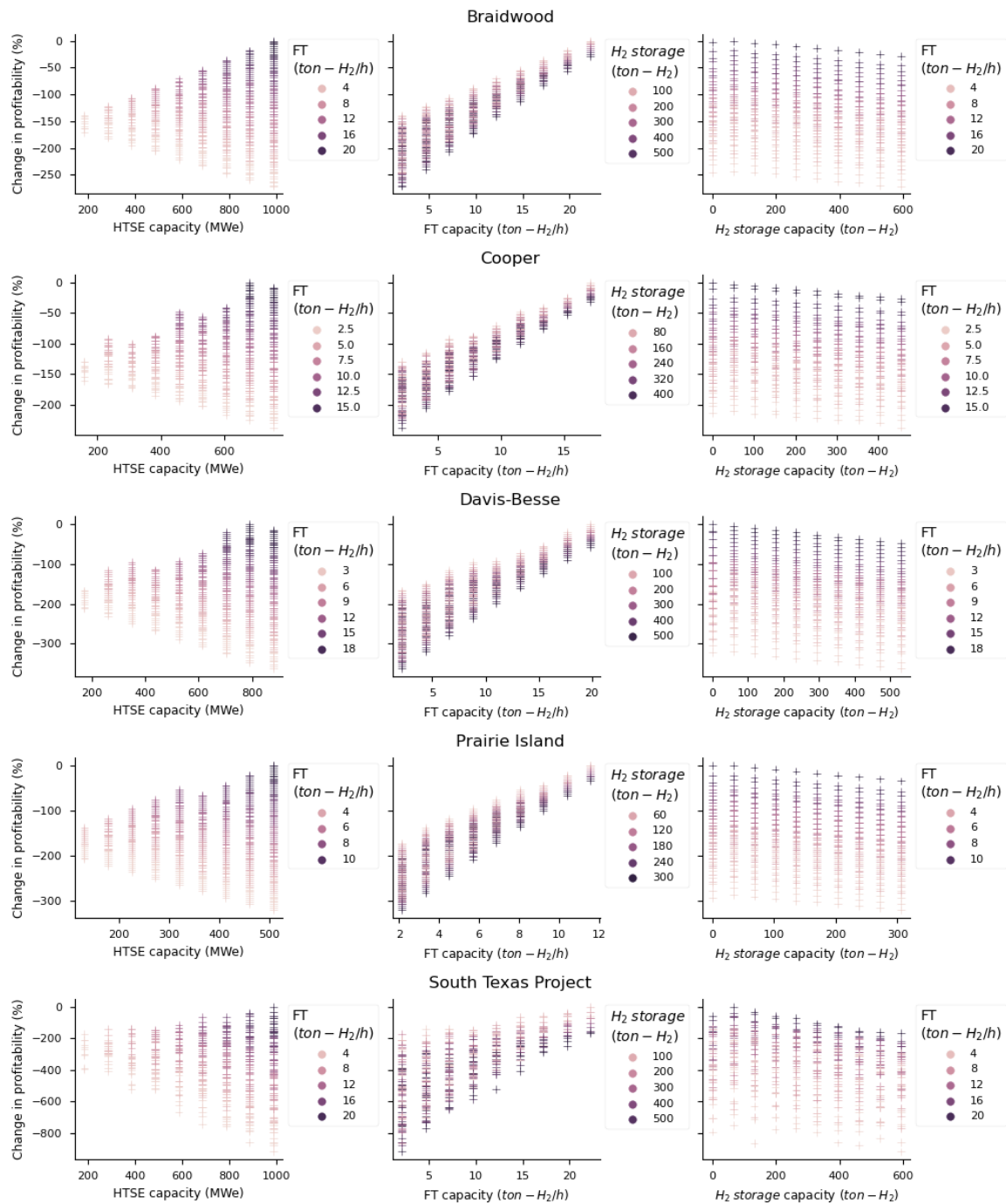

*Figure 7 Configuration Sensitivity Analysis, Profitability as a function of the optimization variables at each location.*

To assess the influence of the configuration of the synfuel IES on its profitability, we use the results of one thousand HERON simulations corresponding to each combination of capacities for the optimized components namely the HTSE, the FT, and the hydrogen storage. These simulations were run to determine the optimal configuration of the synfuel IES at each location. Figure 7 shows the sensitivity of the NPV to the synfuel IES configuration. The capacity of the FT process has the largest effect on the profitability of the synfuel IES, with small reductions in size leading to more than 100% reduction in profitability making the synfuel IES less profitable than the BAU NPP. A reduction in the capacity of this component leads to a decrease in revenues from this tax credit, which has a large influence on profitability as seen in the input parameters sensitivity analysis. Due to the higher price of the carbon feedstock at the South Texas Project location a higher FT capacity does not improve the profitability: the increased feedstock cancels out the additional revenues from the hydrogen PTC. The HTSE sizing has the next largest effect: the synfuel IES is no longer better than the BAU NPP once the HTSE decreases beyond roughly 50% of its optimal capacity. The hydrogen storage capacity has a negative impact on the profitability, particularly noticeable at the South Texas Project location. The optimal storage capacity is the smallest one that enables the FT to receive a constant inflow of hydrogen, increasing its capacity only results in additional CAPEX costs while not increasing the flexibility of the system and thus decreases the overall profitability of the synfuel IES. Finally, we note that the relative sizes of the components matter: at a given FT capacity, decreasing the hydrogen storage is beneficial thanks to the corresponding avoided capital costs and for a given HTSE capacity a higher FT capacity further increases the profitability, which confirm the importance of the FT thanks for the hydrogen PTC.

## 3.3 Synfuel production

| Yearly Normalized production (gal/MW) | Min | Max | Mean | Standard deviation |
|---|---|---|---|---|
| Diesel | 1167 | 1493 | 1378 | 157 |
| Jet Fuel | 1982 | 2536 | 2341 | 267 |
| Naphtha | 1806 | 2310 | 2133 | 243 |

*Table 7 Statistical moments of the yearly normalized production of synfuels at all locations.*

To answer our third research question about the production capacity of the synfuel IES we compute the yearly production of synfuel at each location normalized with respect to the capacity of the NPP. Table 7 presents statistics about the normalized yearly production of synthetic diesel, jet fuel, naphtha. The difference between the production of synfuel at different locations is minimal: the standard deviation of the production of each type of synfuel only represent about 10% of the mean production. Table 6 presented the optimal configuration for each location and showed that the optimal configuration corresponded to maximizing the capacity of the HTSE and close capacities for the FT and HTSE components. The normalized synfuel production are thus similar.

## 4  Discussion

We quantified the added profitability of coupling five existing U.S. NPPs with synfuel production processes into a synfuel IES in comparison to their current business model consisting of solely selling electricity to the grid. We found that, at four out of five NPPs, the synfuel IES would be more profitable. For those, the additional NPV generated stretches from $544M to $975M USD (2020) after accounting for an assumed 10% rate of return on investment on the synfuel production process. Thus, our analysis shows that using nuclear power to produce decarbonized synthetic fuel is a more profitable strategy than the current electricity dedicated one. The existing fleet of nuclear power plants is struggling to maintain its economic profitability and plants are shut down due to operational losses[38]. At the Prairie-Island location, electricity prices are low and the economic potential of the BAU use of the NPP is limited. Adjusted for capacity, the revenues from electricity sales are two to four times lower compared to the other locations. With the synfuel IES, the profitability at this NPP improves more than at other locations. This result shows that the NPPs facing the most unfavorable electricity market conditions have the most to gain from diversifying their range of products via the production of synfuel within a synfuel IES. Furthermore, renewables may continue to depress electricity prices[39], increasing the economic potential of the synfuel IES.

When focusing on the key economic drivers of the production of synfuel via nuclear energy, we found three decisive factors: (i) the revenues from the $H_2$ PTC, (ii) the cost of the $CO_2$ feedstock, and (iii) the variability in electricity prices. When the NPP is located far away from high-quality sources of $CO_2$, the cost of capture and transportation for this feedstock can significantly undermine the profitability of the synfuel IES highlighting the importance of the location of the synfuel IES with regards to high-quality $CO_2$ feedstock sources. The synfuel IES is profitable thanks to the hydrogen PTC from the 2022 Inflation Reduction Act. However, this tax credit can only be for projects starting construction by 2033. Finally, as seen with the South Texas Project location, a high variability in electricity prices will make the profitability of the synfuel IES more sensitive to variations and uncertainties in the system's configurations and in the input parameters. At this location, a larger proportion of the revenues come from electricity sales and the bulk of the electricity produced by the NPP is sent to the grid in periods of high prices. The integration with the grid allows the NPP to continue acting as a grid stabilizer in times of high loads or congestion thanks to the flexibility introduced by hydrogen storage.

The challenge of decarbonizing the transportation sector is substantial: it represented about 25% of the global total energy consumption in 2021[9]. Decarbonized synthetic fuels can readily be produced and used as drop-in fuels: no modifications of the current vehicles fleet or infrastructure would be necessary. The scale of the production rate to attain would require significant energy input. We found that the mean yearly normalized production for diesel, jet fuel and naphtha is respectively 1378 gal/MWe, 2341 gal/MWe, and 2133 gal/MWe. Worldwide 392 GWe of nuclear capacity is operable[40] with the USA, France and China representing more than half the existing nuclear capacity with respectively 95.8, 61.4, and 53.3 GWe. If these three nations were to dedicate their entire nuclear fleet to the production of synthetic fuels the emission of 8.6 MMt-$CO_2$ could be avoided[10] while the conversion of the entire global fleet of nuclear reactors would save 16.1 MMt-$CO_2$ worth of emissions. To attain the goals of the Net Zero by 2050 scenario the emissions from the transportation sector must fall from 8 Gt $CO_2$ (2022 level) to 6 Gt CO2 by 2030, representing an emission reduction target of 250 MMt-$CO_2$ per year. Utilizing the current nuclear fleet for the generation of low-carbon synthetic fuels could contribute towards accomplishing 10% of this target. These calculations highlight substantial disparities in the scale of decarbonization requirements and the existing potential, underscoring the significance of synfuel applications for nuclear power plants worldwide and their market potential.

Although we have assessed the robustness of our results against uncertainties in the system and in the input parameters, our analysis is limited in several ways, in particular (i) the choice of the FT pathway to produce synfuels, and (ii) only including considering the current fleet of NPPs and not more advanced nuclear technologies. While the Fischer-Tropsch process is the most mature and promising technology to produce synfuel, future work could include other synfuel production pathways: the methanol pathway is a close contender in terms of technology readiness, and environmental and economic potential. Its increased operational flexibility could also provide more opportunities for the participation in electricity markets. As to the choice of nuclear reactors, we focused on the current fleet since these reactors could enable the immediate deployment of synfuel IES across the U.S. However, the future of nuclear power in the U.S. will include new designs: higher outlet temperatures and smaller capacities could specifically alter the physical and economic coupling of a nuclear reactor with the synfuel production process. The first reactors of the 4$^{th}$ generation should be built by 2030: under this timeline, if they were integrated into synfuel IES they could still benefit from the hydrogen production tax credit. The 2020 IRA also offers investment tax credit from newly built nuclear and production tax credits for electricity from nuclear energy. Future work could focus on examining and comparing different deployment scenarios of new nuclear, assessing the potential of its integration into a synfuel IES and the economic viability depending on the incentives still available at the time of construction. Our study could then serve as a reference point for economic comparisons.

## 5 Data and Code Availability
Data and input files are available in the following [Github repository](#).

## 6 Acknowledgements
We thank Idaho National Laboratory's Integrated Energy Systems program and the US Department of Energy Office of Nuclear Energy's Nuclear Energy University Program (contract number DE-AC07-05ID14517) for funding.## 7 Author Contributions
M. G., methodology, HERON modeling, analysis, writing -original draft, writing – review& editing, visualization; D. W., HTSE modeling, writing - review& editing; W. J., market study; Q.Z., data collection; M.H-D., $CO_2$ cost modeling; F. J., project administration, supervision; M.C., conceptualization, supervision, writing – review& editing; B.K., conceptualization, supervision, writing - review& editing.

# 8  Declaration of Interests

The authors declare no competing interests.

# 9  Inclusion and diversity

We support inclusive, diverse, and equitable conduct of research.

Appendices

# 10 CO$_2$ feedstock transportation cost model

The quantity of CO$_2$ feedstock required by the synfuel production process depends on its size and it is bounded by how much energy the NPP at each location can produce. Table 8 presents the results of the calculation based on the NPP capacity at each location to determine an upper bound on the CO$_2$ demand for the synfuel production process.

| Nuclear Power Plant | Capacity (MW) | Upper bound on CO2 demand (MT/year) |
|---|---|---|
| Braidwood | 2,354 (2 units) | 2,900,000 |
| Cooper | 769 (1 unit) | 1,000,000 |
| Davis-Besse | 894 (1 unit) | 1,100,000 |
| Prairie-Island | 1,041 (2 units) | 1,300,000 |
| South Texas Project | 1,280 (1 unit) | 1,700,000 |

*Table 8 Upper bound on CO$_2$ demand based on the NPP capacities.*

This analysis considered the following sources for the CO$_2$ feedstock: bioethanol, ammonia, natural gas (power plant), coal (power plant), hydrogen via methane reforming, iron/steel, and cement plants. Figure 8 shows the sources and NPPs geographical distribution across the U.S. The volumetric concentration of CO$_2$ in waste-streams is the main driver for the capture cost at the plant, it is presented in Table 9.

| Source | U.S. CO$_2$ emissions (MMT/year) | Concentration CO$_2$ (%vol) |
|---|---|---|
| Bioethanol | 31 | 99.8 |
| Ammonia | 35 | 97.1 |
| Natural Gas | 54 | 99 |
| Hydrogen | 44 | 44.5 |
| Iron/Steel | 72 | 23.2-26.4 |
| Cement | 67 | 22.4 |

*Table 9 CO$_2$ volumetric concentration by source[41].*

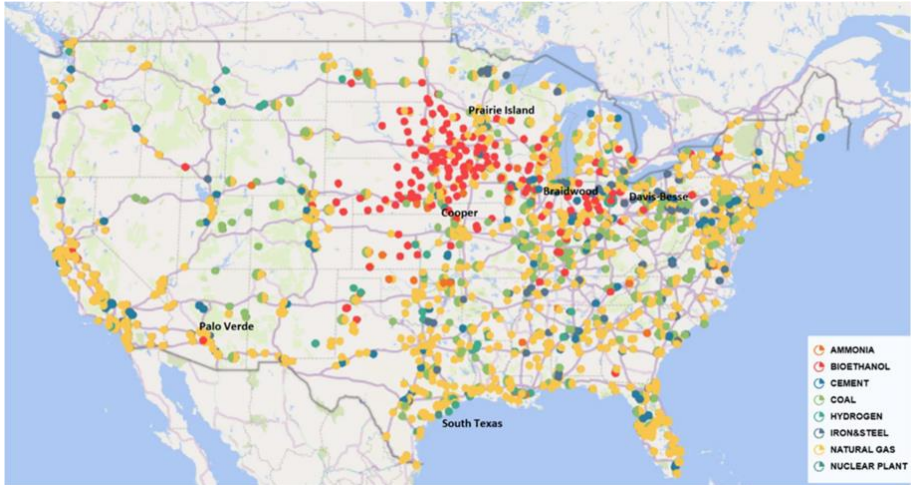

*Figure 8 Geographical distribution of $CO_2$ sources and NPPs[33].*

The cost of $CO_2$ capture and compression for each type of source was computed following a procedure developed by NREL[34]. High purity sources do not require a capture process to meet pipeline requirements while for lower purity streams additional costs come from a methyl diethanolamine-acid gas removal process. Figure 9 presents the results of the compression costs calculation for each type of source.

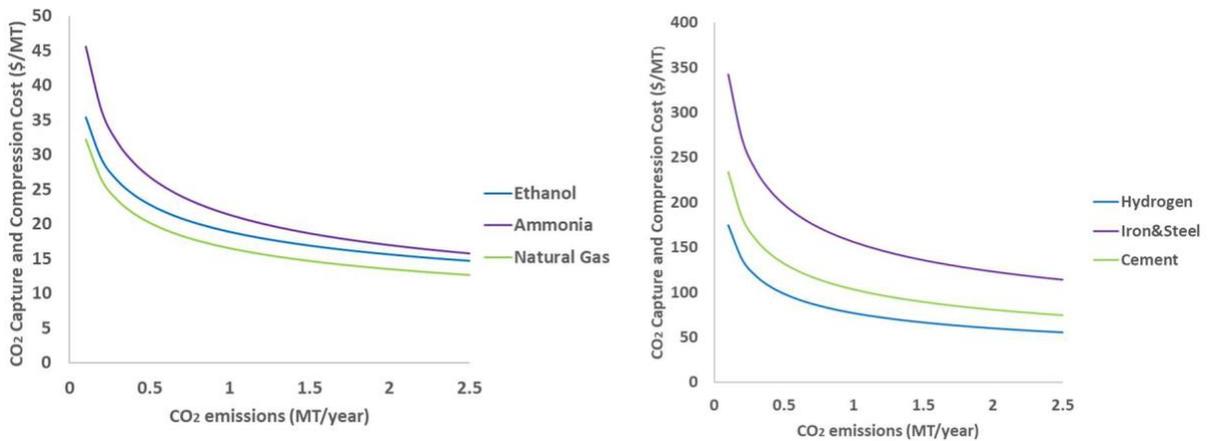

*Figure 9 Capture and compression cost by source.*

Once captured the CO2 feedstock is transported via pipelines from the sources to the synfuel IES location. A supply curve for each location is built by identifying the set of the closest sources presenting a sufficient cumulative production capacity to satisfy the upper bound determined at the beginning of this analysis. The National Energy Technology Laboratory transport model[42] is then used to compute the cost of transportation. The following inputs for this model were used:
- The length of the pipeline is computed via the Open Source Routing Machine.
- The annual flow rate is adjusted to represent the source considered.
- The region of the NPP
- The electricity cost in the region in which the pipeline is located.

The $CO_2$ capture cost computed from the information presented in Table 9.

The model enables the computation of the optimal pipeline diameter and number of booster pumps to minimize the transportation cost. The cumulative average capture and transportation costs can then be computed to build the supply curve for each NPP location:

$$\text{Cumulative average capture and transportation costs} = \frac{\sum_{i \in N} c_i q_i}{\sum_{i \in N} q_i}$$

Where $N$ is the set of sources considered, $c_i$ is the cost of capture and transportation for each source, $q_i$ the amount of $CO2$ transported from each source.

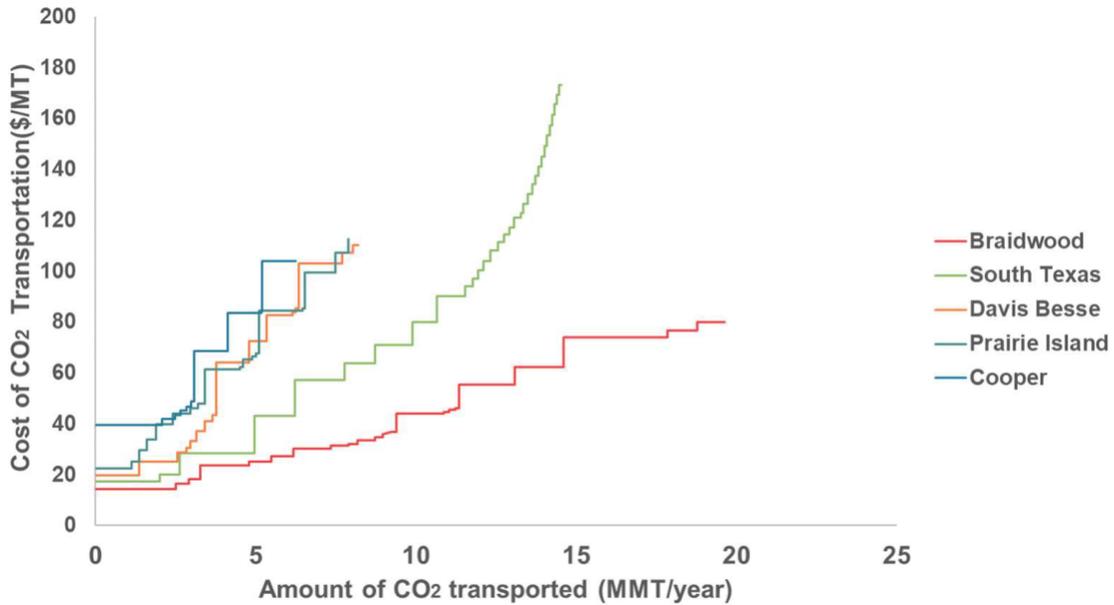

*Figure 10 $CO_2$ supply curve for each location presenting the cost of the $CO_2$ feedstock as a function of the amount of $CO_2$ transported.*

# 11 Synthetic Time Series via the Auto-Regressive Moving Average model

A Fourier series decomposition can describe the periodicity in electricity prices. These periodic patterns can occur every year, season, month, … Fourier modes, as defined by the following equation are identified, and subsequently subtracted from the price signal:

$$F = \sum_c \sum_{i=0}^{k} a_i sin\left(\frac{2\pi i}{c} t\right) + b_i cos\left(\frac{2\pi i}{c} t\right)$$

Where $c$ is the set of characteristic time periods for which the original signal presents a periodicity, $f$ are the frequencies subdividing the characteristic time lengths, and $a$ and $b$ are computed to fit $F$ to the original signal.

The pricing data is detrended via a Fourier analysis. The remaining noise is standardized to a standard normal distribution and the residual normalized signal is modelled thanks to the ARMA model via the ARMA model via the following equation:

$$y_t = \sum_i^P \phi_i y_{t-i} + \epsilon_t + \sum_j^Q \theta_j \epsilon_{t-j}$$

The first term corresponds to the auto-regressive term, P being the maximum number of auto-regressive terms. Then in the moving average term, Q is the maximum number of moving average terms. $\epsilon$ corresponds to Gaussian noise; φ and θ are fitted to maximize likelihood.

Here are presented the mathematical foundations of the Fourier and ARMA signal processing methods; the implementation of the Fourier decomposition and the ARMA in the RAVEN code will provide even further details[28]. The synthetic time series obtained via Fourier and ARMA can be used to reproduce the variability of historical electricity prices data, then applied for future years, as the prediction of electricity prices is outside the scope of our analysis.

## 11.1 ARMA Model Validation

|  | Braidwood | | Cooper | | Davis-Besse | | Prairie-Island | | South Texas Project | |
|---|---|---|---|---|---|---|---|---|---|---|
|  | **Historical** | **Synthetic** | **Historical** | **Synthetic** | **Historical** | **Synthetic** | **Historical** | **Synthetic** | **Historical** | **Synthetic** |
| mean | 33.0 | 33.0 | 25.8 | 25.5 | 33.0 | 33.0 | 9.4 | 9.4 | 41.2 | 41.6 |
| std | 23.1 | 23.1 | 110.3 | 110.6 | 23.1 | 23.1 | 15.1 | 15.1 | 290.7 | 290.5 |
| min | 2.3 | 2.3 | -65.1 | -65.1 | 2.3 | 2.3 | -66.2 | -66.2 | -20.2 | -20.2 |
| 25% | 22.7 | 22.6 | 14.8 | 14.4 | 22.7 | 22.6 | -0.5 | -0.6 | 18.9 | 16.8 |
| 50% | 28.8 | 28.8 | 19.3 | 19.2 | 28.8 | 28.8 | 0.5 | 0.6 | 23.9 | 24.4 |
| 75% | 37.0 | 37.1 | 27.2 | 27.2 | 37.0 | 37.1 | 21.5 | 21.5 | 32.0 | 36.4 |
| max | 933.7 | 933.7 | 4231.0 | 4231.0 | 933.7 | 933.7 | 97.0 | 97.0 | 8996.8 | 8996.8 |
| kurtosis | 243.3 | 241.8 | 755.9 | 750.0 | 243.3 | 241.2 | 1.2 | 1.2 | 699.6 | 696.7 |
| skewness | 10.9 | 10.8 | 26.8 | 26.6 | 10.9 | 10.8 | 1.3 | 1.3 | 25.8 | 25.7 |

*Table 10 Synthetic and historical electricity prices distribution comparison for the validation of ARMA models.*

To perform the validation of the ARMA model trained on historical electricity market data we compute and compare statistical moments for the historical and synthetic data at each location (Table 10). The synthetic data should reproduce the historical data and we observe that the median, quartiles, minimum and maximum are identical. The kurtosis and the skewness of the distribution are close. They are respectively a measure of the importance of outliers in the distribution and of the symmetry of the distribution and these characteristics of the electricity prices distribution will influence the behavior of the synfuel IES as extreme electricity prices will change the relative value of the synfuel products compared to electricity.

# 12 Synfuel Market and Prices Forecasts

According to the U.S. Energy Information Administration (EIA), 88% of U.S. transportation energy comes from gasoline, distillates, and jet fuel. We analyze the market for these fuels with the addition of naphtha to explore market opportunities for synthetic fuel. Gasoline is not being evaluated for production using the FT plant. Gasoline market statistics are used as reference for naphtha since it is used as a blendstock for gasoline production. Regional fuel trends are analyzed to understand market opportunities for specific NPPs located in five unique U.S. Census Bureau regions. Pricing changes on a regional basis due to variations in supply and demand. State-specific fuel taxes are applied.

Fuel use statistics for diesel, gasoline, and jet fuel were taken from the EIA Annual Energy Outlook. No data is available from EIA for naphtha. Historical price correlation between naphtha and gasoline were used to forecast naphtha prices. In the U.S., we present the fuel use volumes to understand market characteristics. According to EIA forecasts, and accounting for all fuel uses, 60.4 billion gallons of distillate fuel oil (diesel) was used during 2022. The transportation sector uses 128 million gallons of diesel each day. Thanks to its high energy density and reduced flammability, diesel is the preferred source of fuel for many applications. By comparison, 133 billion gallons of gasoline and 24 billion gallons of jet fuel were used in 2022. Regional market size fluctuates based on the regional industrial mix and population size. Figure 11 shows a map of the U.S. Census regions used by EIA for price and energy use statistics and in our analysis for modeling production scenarios. Table 11 presents the volume of fuel used by region in the U.S. in 2022.

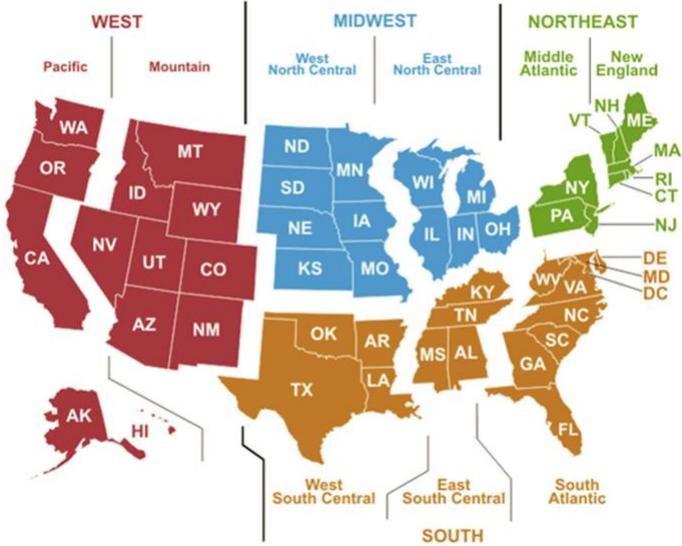

*Figure 11 Map of the U.S. census regions[35].*

| U.S. Fuel Market by Region and Volume (billions of gallons, 2022) | | | |
|---|---|---|---|
| Region | Distillate Fuel Oil | Jet Fuel | Gasoline |
| United States | 60.4 | 23.8 | 132.8 |
| West North Central | 6.5 | 0.7 | 9.6 |
| Pacific | 6.3 | 6.1 | 19.0 |
| East North Central | 8.1 | 2.1 | 18.7 |
| West South Central | 11.4 | 2.6 | 18.9 |

*Table 11 U.S. fuel market by region and volume, Source: EIA, Annual Energy Outlook 2022.*

EIA provides long-term energy use forecasts under multiple scenarios. For this analysis the standard reference scenario was used. Each of the NPP in our analysis is in a census region identified in Table 5. Fuel price forecasts from the EIA were adjusted by removing applicable federal and state fuel taxes using the most current data (August 2022). Costs associated with marketing, distribution and transportation were calculated and subtracted from EIA price forecasts to reduce retail prices to levels equivalent to refinery gate prices as presented in Table 12.

| Retail Fuel Price Adjustment Table | | | | | | |
|---|---|---|---|---|---|---|
| State | Illinois | Minnesota | Ohio | Texas | Nebraska | California |
| Region | East North Central - IL | West North Central - MN | East North Central - OH | West South Central | West North Central - NE | Pacific |
| Nuclear Plant | Braidwood | Prairie Island | Davis-Besse | South Texas | Cooper | |
| Transport Method | Rail | Water | Water | Water | Water | Rail |
| **State+Federal Tax ($/gal) (Aug, 2022)** | | | | | | |
| Diesel Fuel | $0.952 | $0.530 | $0.714 | $0.444 | $0.495 | $1.146 |
| Jet Fuel | $0.219 | $0.369 | $0.219 | $0.219 | $0.249 | $0.239 |
| Gasoline | $0.817 | $0.470 | $0.569 | $0.384 | $0.441 | $0.835 |
| Naphtha | $0.136 | $0.136 | $0.136 | $0.136 | $0.136 | $0.136 |
| **Marketing + Distribution (% of Retail Price, 2021)** | | | | | | |
| Diesel Fuel | 20.2% | 20.2% | 20.2% | 20.2% | 20.2% | 20.2% |
| Gasoline | 15.6% | 15.6% | 15.6% | 15.6% | 15.6% | 15.6% |
| **Marketing ($/gal)** | | | | | | |
| Naphtha | $0.06 | $0.06 | $0.06 | $0.06 | $0.06 | $0.06 |
| Jet Fuel | $0.06 | $0.06 | $0.06 | $0.06 | $0.06 | $0.06 |
| **Distribution ($/gal) (Based on Transport Method)** | | | | | | |
| Jet Fuel (Pipeline) | $0.12 | $0.12 | $0.12 | $0.12 | $0.12 | $0.12 |
| Naphtha | $0.36 | $0.03 | $0.03 | $0.03 | $0.03 | $0.36 |

Sources: The Energy Journal, U.S. Dept. of Transportation, EIA

*Table 12 Regional fuel price adjustment factors.*

Price forecasts from EIA were not available for naphtha, but historical prices were available from other sources[43]. Naphtha and gasoline historical prices are highly correlated. Therefore, EIA regional gasoline price forecasts for gasoline were used to compute forecasts for regional naphtha prices. Figure 12 provides an example of a comparison for the fluctuations of regional distillate fuel oil prices over time.

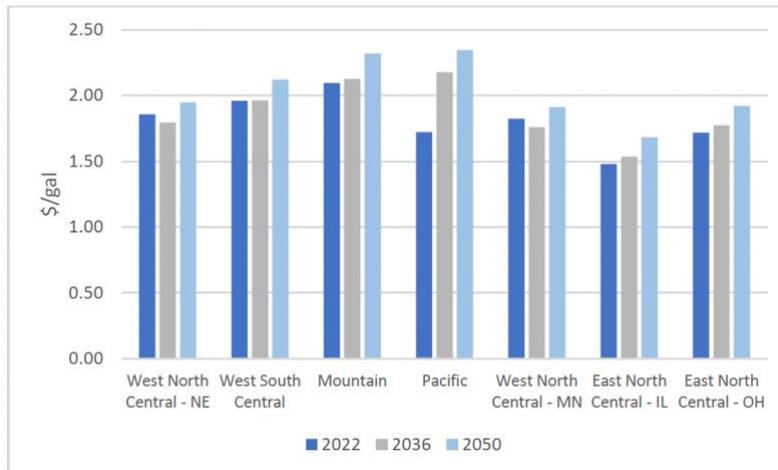

*Figure 12 Evolution of adjusted distillate fuel oil prices across U.S. census regions.*

# 13 Sensitivity Analysis Results

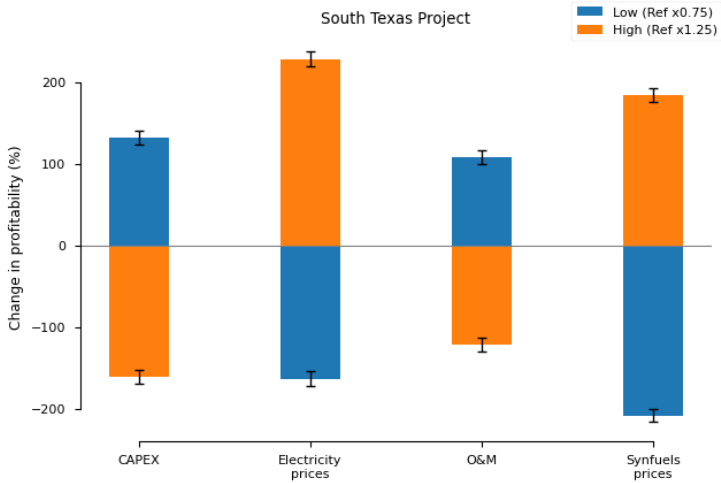

*Figure 13 The perturbation of input parameters has a significant influence on the synfuel IES profitability at the South Texas Project location, Sensitivity analysis results at the South Texas Project location for the CAPEX, O&M costs, electricity prices and synfuel prices.*

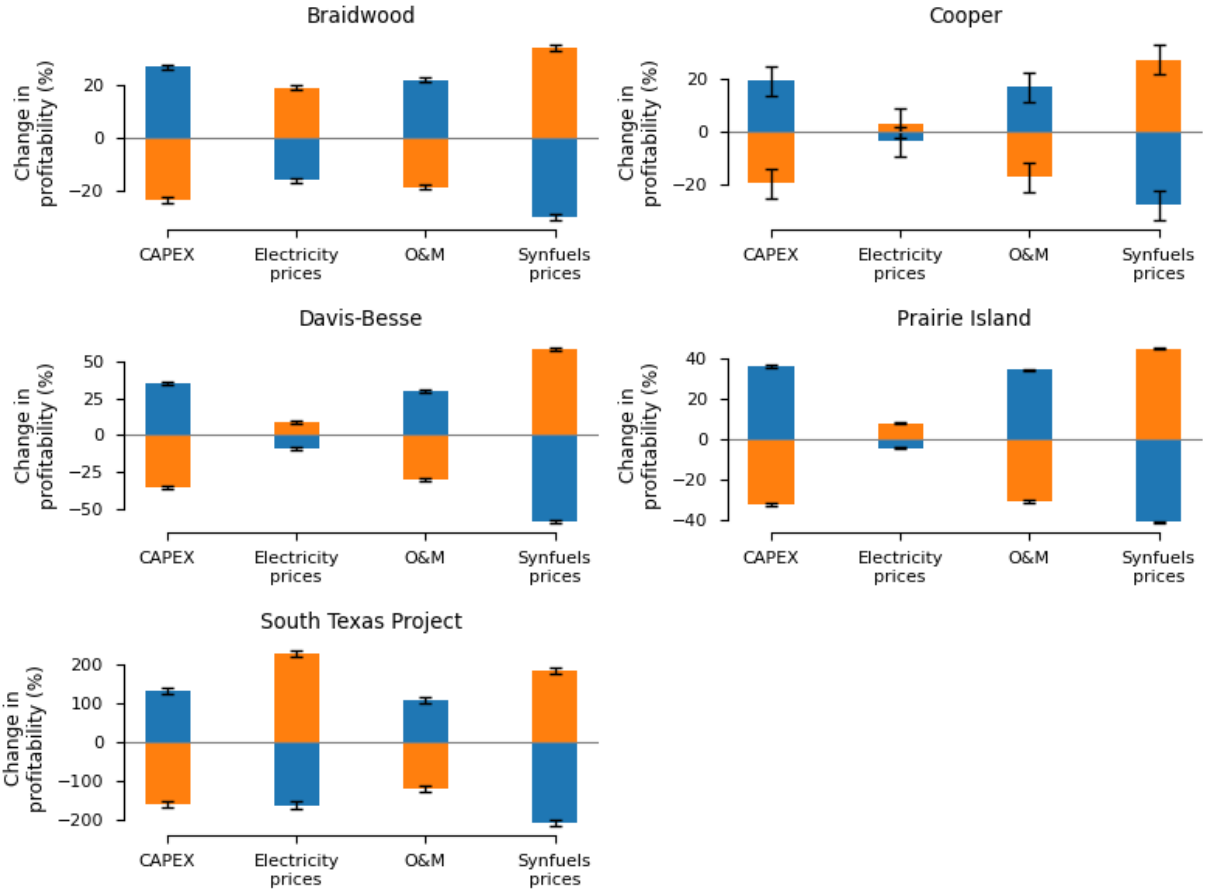

*Figure 14 Sensitivity analysis results for the CAPEX, O&M costs, electricity prices, and synfuels prices at all locations.*

## 14 The grid stabilizer role of the NPP within the synfuel IES

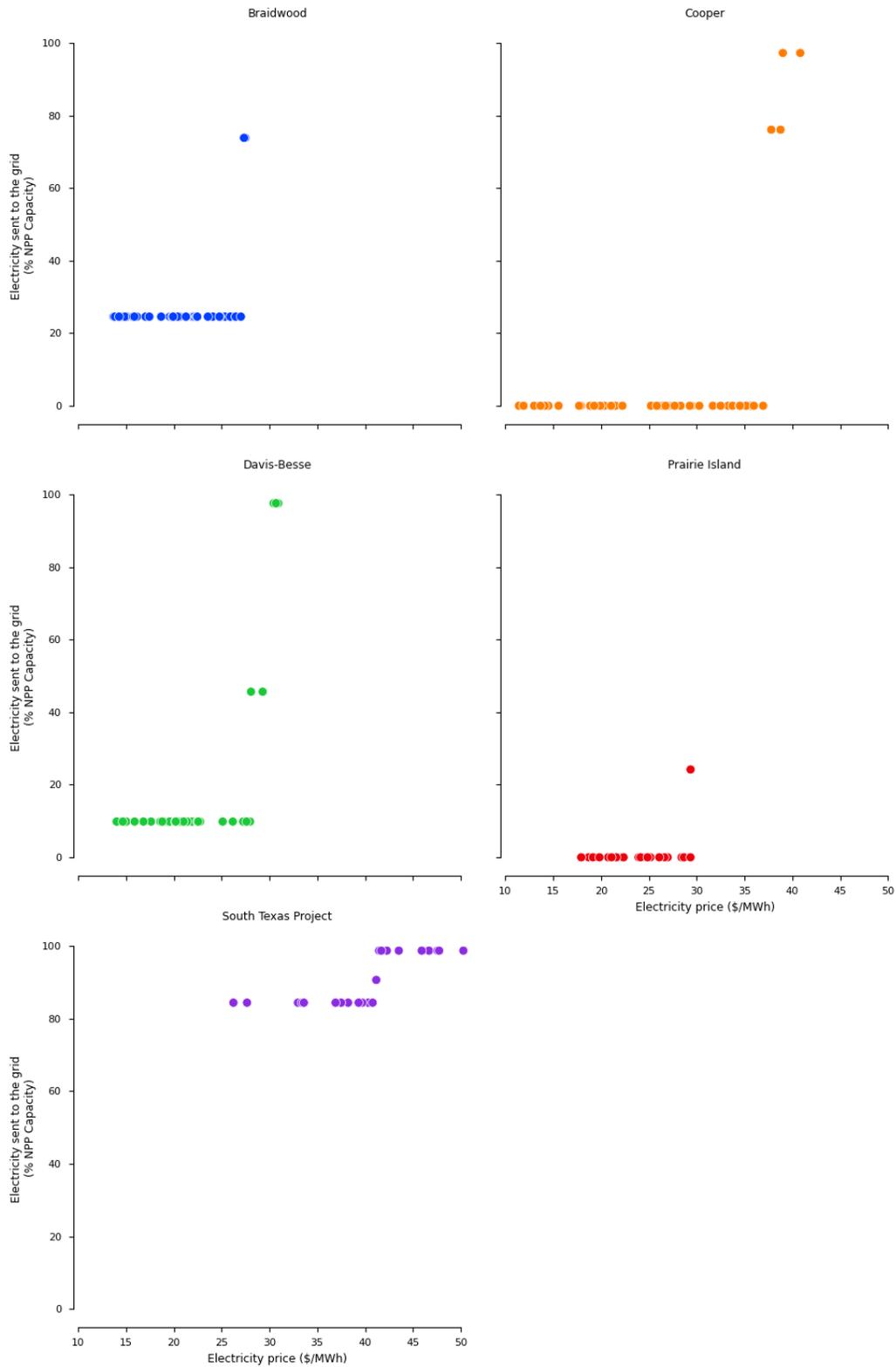

*Figure 15 Participation of the synfuel IES in the electricity market at each location, Year 2019.*

**Error! Reference source not found.** Figure 15 presents the energy sent to the grid as a function of the electricity price for each location during the year 2019. We observe the existence of an electricity price threshold at each location: a certain electricity price prompts the IES to sharply increase the proportion of electricity produced by the NPP sent to the grid. At the Prairie-Island location, electricity prices are on average low (see Figure 3) and thus most of the time the portion of the NPP power sent to the grid is low. On the other hand, at the Braidwood location, electricity prices are higher on average, and thus more power is sent to the grid. Examining $CO_2$ feedstock costs for those two locations (Figure 10), we note that they are lower for the Braidwood location and thus the synfuel IES will start sending more electricity to the grid at a higher threshold electricity price compared to the Prairie-Island location. Therefore, depending on the economics of the synfuel production process inflows (price of $CO_2$ feedstock), the NPP supports the stability of the grid during periods of high prices which are often correlated with high demand or grid congestion.